\documentclass[11pt,a4paper]{article}                                                     

\newtheorem{lemma}{Lemma}
\newtheorem{theorem}{Theorem}

\newtheorem{remark}{Remark}

\usepackage{amsmath}
\usepackage{easybmat}
\usepackage{graphicx}
\usepackage{subfigure}
\usepackage{cite}
\def \QED{\mbox{\rule[0pt]{0.7em}{0.7em}}}

\title{\bf Distributed Stochastic Approximation: Weak Convergence and Network Design}

\author{Milo\v{s} S. Stankovi\'{c}\thanks{Innovation Center, School of Electrical Engineering, University of Belgrade,
Serbia (e-mail: milos.stankovic@ic.etf.rs).}, \hspace{2mm}Nemanja Ili\'{c}\thanks{School of Electrical Engineering, University of Belgrade, Serbia (e-mail: {nemili@etf.rs}).}\hspace{2mm} and \\ Srdjan S. Stankovi\'{c}\thanks{School of Electrical Engineering, University of Belgrade, Serbia, and the Vlatacom Institute, Belgrade, Serbia (e-mail: stankovic@etf.rs).} }

\date{}

\begin{document}

\maketitle
\let\thefootnote\relax\footnote{This work was supported by the EU Marie Curie CIG (PCIG12-GA-2012-334098).}

\begin{abstract}
This paper studies distributed stochastic approximation algorithms
based on broadcast gossip on communication networks represented by
digraphs. Weak convergence of these algorithms is proved, and an
associated ordinary differential equation (ODE) is formulated
connecting convergence points with local objective functions and
network properties. Using these results, a methodology is proposed for
network design, aimed at achieving the desired asymptotic behavior at
consensus. Convergence rate of the algorithm is also analyzed and
further improved using an attached stochastic differential equation.
Simulation results illustrate the theoretical concepts.
\end{abstract}

\section{Introduction}
\emph{Stochastic approximation} (SA) algorithms have been in the focus of researchers for more than sixty years,
\emph{e.g.}, \cite{hfchen,bmp,ky} and the references therein. More than thirty years ago a \emph{distributed} SA
algorithm (DSA) based on a first-order \emph{consensus} scheme was proposed in \cite{tsi}, and later analyzed in
detail for both constant and tapering step-sizes \cite{tsi,ber1,tba,ky1,ky2}. In the context of recently growing
interest for distributed algorithms, networked control systems and multi-agent systems, new contributions to the
DSA algorithms have appeared, motivated by numerous applications in optimization, estimation, detection and control
in which decentralization or distribution of functions is either dictated by information structure constraints, or
aimed at parallelizing computations and increasing both robustness and scalability \cite{nop,no,rnv,ned,blo,bfh,bij,smdieee1}.
The basic problem setting in all these approaches follows essentially the main line of thought from \cite{tsi,tba,ky1}.
Depending on the context, specific assumptions have been adopted, especially in relation with the inter-agent
communications. Communication links have been adopted to be either constant (\textit{e.g}., \cite{ts}) or random
(\textit{e.g.}, \cite{no,nop,rnv,lo,huang}), assuming double stochasticity of the
\emph{random communication (gossip) matrices}. In \cite{ned,bij,bfh,mbf},
this assumption has been somewhat relaxed, assuming double stochasticity of the mean value of these matrices. In \cite{mbf} an analysis is provided showing that non-doubly stochastic matrices generally influence the limit points but without their precise characterization in terms of network properties. In \cite{neol} this assumption has been drastically relaxed, requiring, however, network uniform connectivity and the knowledge of the number of out-neighbors at each time instance (which, in general, requires certain bidirectional communications).
 \par
In this paper, we approach DSA algorithms from a general standpoint, assuming random communications within a \emph{directed graph}. More specifically, we consider two structurally distinct general
representatives of
DSA algorithms based on \emph{broadcast gossip}, differing by the order of updating and convexifying local parameter
estimates (see, \emph{e.g.}, \cite{aysal,ned,ts}). Starting from a preliminary analysis related to the broadcast
gossip scheme itself, \emph{weak convergence} of both DSA algorithms is analyzed using the methodology presented by
Kushner
and Yin in \cite{ky1} in relation with only one specific DSA algorithm. The weak convergence methodology has been
adopted, starting from the following general arguments: 1) when working with practical systems, gain parameters
almost never go to zero; 2) one can locate the points in the neighborhood of which the processes spend
asymptotically most of the time (by defining an associated ordinary differential equation -- ODE); 3) assumptions about
the noise are generally weaker than those in the with probability one (w.p.1) methods;
4) it is much easier than in the case of w.p.1 methods to come to valid conclusions for practice --
often by direct inspection  of the related conditions (see, \textit{e.g.}, \cite{ky,ky1} and the references therein). The first contribution of the paper is the proof of weak convergence of the analyzed algorithms and derivation of an associated ODE
defining \emph{location of the convergence points} in terms of local objective
functions and network parameters.
Based on these results, the second contribution of the paper is the proposal of \emph{two algorithms for network
parameter design} aimed at obtaining the desired convergence points; the existence of solution has been proved for these algorithms. Needs for such design methods are present in all domains of application of DSA algorithms:
it is possible, \emph{e.g.}, to handle directly the problem of distributed optimization of criterion functions given in the form of a
weighted sum of local criterion functions.
As the third contribution, it is shown that the normalized asymptotic error can be modeled by a
\emph{stochastic differential equation} (SDE) which offers a possibility to \emph{improve} \emph{the convergence rate} of the algorithms. The given simulation results illustrate the main theoretical
conclusions and give some additional practical insights.
\section{Problem formulation and algorithms}
Consider a network of $N$ agents, represented by a directed graph ${\cal G}=({\cal N},{\cal E})$, where
${\cal N}=\{1, \ldots, N\}$ is the set of nodes, and $ {\cal E}= \{ (i,j) \}$ the set of directed links from node $i$ to
node $j$. Let $A_{{\cal G}}$ be the adjacency matrix of ${\cal G}$. Define the sets of out-neighbors and in-neighbors of
the
$i$-th node ${\cal N}^{o}_{i}=\{j \in {\cal N} | (i,j) \in {\cal E} \}$ and ${\cal N}^{i}_{i}=\{j \in {\cal N} | (j,i)
 \in {\cal E} \}$, respectively. Assume that each agent has an internal clock that ticks independently from the other
  clocks according to a rate $\mu_{i} > 0$ Poisson process, $i=1, \ldots, N$.
At each tick of the internal clock, the node $i$ broadcasts its current (or updated) state to the nodes
$j \in  {\cal N}^{o}_{i}$; due to possible link failures, some neighbors may not receive this message.
When successfully received, the message leads to an \emph{update of the state of the corresponding node}, while the remaining nodes preserve their previous states. Such a type of communication scheme belongs to the class of \emph{broadcast gossip algorithms}, see \emph{e.g.} \cite{aysal,boydgossip} and the references therein. Following \cite{ned}, we replace in our analysis the set of local clocks by a \emph{global virtual clock} with the rate $\mu=\sum_{i} \mu_{i}$ that ticks whenever any local clock ticks; the $n$-th
tick of the global clock is considered as the discrete time instant (iteration) $n$. At any instant $n$, let $p_{i} > 0$ be the probability for $i$-th clock to tick,  ${\cal J}_{i}(n) \subseteq  {\cal N}^{o}_{i} $ the set of nodes that received the message from node $i$, $p_{ij} ={\rm P} \{ j \in {\cal J}_{i}(n) \}>0 $ for $j\in {\cal N}^{o}_{i}$ the probability of $j$-th node to receive the message from node $i$ and $\bar{d}_{j}=\sum_{i \in  {\cal N}^{i}_{i}}p_{ij}p_{i} $ the probability of $j$-th node to receive the message from an in-neighbor and update its state.
\par
\emph{Let the $i$-th clock ticks at the instant} $n$ and let $X^{j}_{n} \in R^{p}$ be the current state of the $j$-th agent, $j=1, \ldots, N$. In general, the agents are supposed to generate their new states in the following two distinct ways:
\par
\emph{Algorithm update-convexify (AUC)}. At the \emph{first step}, the agent $i$ calculates its \emph{updated} state
\begin{equation} \label{auc1}
\hat{X}^{i}_{n+1}=X^{i}_{n}+ \varepsilon F^{i}(X_{n}^{i}, \xi_{n}^{i}),
\end{equation}
and sends it to its neighbors, where $F^{i}(X_{n}^{i}, \xi_{n}^{i})$ is the ``observation" of agent $i$ at time $n$,  $\xi_{n}^{i}$ represents local uncertainty and $\varepsilon > 0$ the step-size.
The neighbors $j \in {\cal J}_{i}(n)$ generate, at the \emph{first step}, their own updated states $\hat{X}^{j}_{n+1}$ using (\ref{auc1}),
and calculate, at the \emph{second step}, their new states as \emph{convex combinations}
\begin{equation} \label{auc2}
X^{j}_{n+1}= \beta_{ij} \hat{X}^{j}_{n+1}+ (1- \beta_{ij}) \hat{X}_{n+1}^{i},
\end{equation}
where $0 < \beta_{ij} < 1$, while $X^{i}_{n+1}=  \hat{X}^{i}_{n+1}$ and  $X^{k}_{n+1}=  X^{k}_{n}$ for $k \notin {\cal J}_{i}(n)$, $k \neq i$.
\par
\emph{Algorithm convexify-update (ACU)}. The agent $i$ sends at instant $n$ its current state $X_{n}^{i}$ to the neighboring nodes. At the \emph{first step}, the agents  $j \in {\cal J}_{i}(n)$  calculate \emph{convex combinations}
\begin{equation} \label{acu1}
\hat{X}^{j}_{n}= \beta_{ij} X^{j}_{n}+ (1- \beta_{ij}) X_{n}^{i},
\end{equation}
 and, at the \emph{second step}, update their states using
\begin{equation} \label{acu2}
X^{j}_{n+1}=\hat{X}^{j}_{n}+ \varepsilon F^{j}(\hat{X}_{n}^{j}, \xi_{n}^{j});
\end{equation}
$X^{k}_{n+1}=  X^{k}_{n}$ for all $k \notin {\cal J}_{i}(n)$.
\par
AUC has been described in \cite{smdauto,smdieee1,bfh,bij}, and ACU  in \cite{ned}; in \cite{ts} both schemes are presented in a deterministic context, while the algorithm analyzed in \cite{ky1} can be considered as a special case of ACU (see the analysis given below).
\par
Define $X_{n}=(X^{1}_{n}, \cdots, X^{N}_{n})$ (the notation $(x_1, \ldots, x_N) $ is used throughout the paper to represent the column vector obtained by concatenating vectors $x_1, \ldots, x_N$), $\tilde{A}_{n}=A_{n} \otimes I_{p}$,  where  $A_{n}=[a_{jk}(n)]$ is a random matrix defined in such a way that
  \begin{equation} \label{aij}
  a_{jk}(n)= \left\{ \begin{array}{l} 1, \;\;\; j \notin \mathcal{\mathcal{J}}_{i}(n), k=j \\ \beta_{ij}, \;\;\; j \in \mathcal{J}_{i}(n), k=j \\ 1-\beta_{ij}, \;\;\; j \in \mathcal{J}_{i}(n), k=i \\ 0 \;\;\; {\rm elsewhere} \end{array} \right.
  \end{equation}
for $i,j,k=1, \ldots, N$. Introducing $Y_{n}=(Y^{1}_{n}, \cdots, Y^{N}_{n})$, $\xi_{n}=(\xi^{1}_{n}, \cdots, \xi^{N}_{n})$ and $F(Y_{n}, \xi_{n})=(F^{1}(Y_{n}^{1}, \xi_{n}^{1}), \cdots, F^{N}(Y_{n}^{N}, \xi_{n}^{N}))$, we obtain from (\ref{auc1})--(\ref{acu2}) the following global
representation  for both AUC and ACU \emph{at the network level}:
\begin{equation} \label{alg}
X_{n+1}=\tilde{A}_{n} X_{n}+ \varepsilon \tilde{C}_{n} F(Y_{n}, \xi_{n})
\end{equation}
where:

\emph{In the case of AUC:}  $\tilde{C}_{n}$ $=A_{n}D_{n} \otimes I_{p}$, with $D_{n}={\rm diag} \{d_{1}(n), \cdots, d_{N}(n) \}$, where $d_{k}(n)$
 is a binary random variable equal to one if $k \in {\cal J}_{i}(n) \cup \{i \}$ and zero otherwise, reflecting communication dropouts, and  $Y_{n}=X_{n}$;

\emph{In the case of ACU:}   $\tilde{C}_{n}=D^{0}_{n} \otimes I_{p}$, where $D_{n}^{0}$ has the same elements as $D_{n}$, except $d_{i}(n)$ which is equal to zero, and $Y_{n}=\tilde{A}_{n} X_{n}$.

Let $\{ \mathcal{F}_{n} \}$ be an increasing sequence of $\sigma$-algebras such that $ \mathcal{F}_{n}$ measures $\{X_{i}, i \leq n; \xi_{i}, A_{i}, i < n\}$ and let $E_{k}\{ \cdot \}$ denote $E \{ \cdot| \mathcal{F}_{k} \}$.
\section{Convergence Analysis}
\subsection{Broadcast Gossip Communication Scheme}
In order to proceed with the analysis of (\ref{alg}), we need a slight generalization of Lemma 2.1 in \cite{ky1} (which treats the case of fully connected graphs).
\par
(A.1) The digraph ${\cal G}$ is strongly connected.
\par
Define $\Phi(n|k)=A_{n} \cdots A_{k}$  for $n \geq k$, with $\Phi(n|n+1)=I_{N}$. Therefore,
$E \{ \Phi(n|k)\}$= $\bar{A}^{n-k+1}$, where $\bar{A}= E\{A_{n}\}=\sum_{l} A^{[l]} \pi_{l}$,
$A^{[l]}$ denoting
 $l$-th realization of $A_{n}$ and $\pi_{l}$ its probability. According to (A.1), $\bar{A}$ is primitive, so that
 $\bar{\Phi}= \lim_{n \to \infty} E \{ \Phi(n|k) \}=  \lim_{n \to \infty} \bar{A}^{n-k+1}$ exists for any fixed $k$
 and has the form $\bar{\Phi}= \left[ \begin{BMAT}{ccc}{c} \bar{\phi}^{T} & \cdots & \bar{\phi}^{T}
 \end{BMAT} \right]^{T}$, where  $\bar{\phi}=(\bar{\phi}_{1}, \cdots, \bar{\phi}_{N})^{T}$,
 $\bar{\phi}_{i} > 0$, $i=1, \ldots, N$; moreover, $\bar{\phi} \bar{A}=\bar{\phi}$. Notice that the same
 broadcast gossip scheme is analyzed in \cite{aysal}, assuming \emph{a priori} that $p_{i}=p_{j}=1/N$,
 $\beta_{ij}=\beta$, $i,j=1, \ldots, N$ and $\mathbf{1}^{T} \bar{A}=\mathbf{1}^{T}$; then,
 $\bar{\phi}= \frac{1}{N} \mathbf{1}^{T}$ ($\textbf{1}=(1, \cdots, 1)$).

\begin{lemma} Let (A.1) hold. Then:
\par
a) $\Phi_{k}=\lim_{n \to \infty} \Phi(n|k)$ exists w.p.1 and its rows are all equal, \emph{i.e.}, $\Phi_{k}= \left[ \begin{BMAT}{ccc}{c} \phi(k)^{T} & \cdots & \phi(k)^{T}  \end{BMAT} \right]^{T} $, with $\phi(k)=( \phi_{1}(k), \cdots, \phi_{N}(k) )^{T} $;
\par
b) $E \{ |\Phi(n|k)-\Phi_{k}| \} \rightarrow 0$ geometrically as $n-k \rightarrow \infty$, uniformly in $k$ ($|\cdot|$ denotes the infinity norm). \end{lemma}

{\bf Proof:}
For the adopted broadcast gossip scheme, there exists a scalar $p_{0} > 0$ such that for all $n$ the probability that a node $i$ communicates to any node $j \in {\cal N}^{o}_{i} $ is greater than or equal to $p_{0}$, for all $i$. By (A.1), there exist $\alpha' >0$ and  $m' > 0$ such that for all $n$ \emph{all the elements} of  $\Phi(n+m'|n)$ are $\geq \alpha'$ with some positive probability $p_1$; this implies that
 there exist $\alpha'' > 0$ and an increasing sequence of finite w.p.1 random times $\{\nu_{i}\}$ such that the components of $\Phi(\nu_{i+1}|\nu_{i})$ are all $\geq \alpha''$ w.p.1. Now, the result follows from Lemma~~2.1 in \cite{ky1} and Lemma~5.2.1 in \cite{tsi}. \hspace*{\fill}\QED

\begin{remark} Notice that assertion a) can be derived directly using the recently published general results \cite{tn1,tn2,hts} covering a wide class of gossip protocols,
and providing necessary and sufficient conditions for convergence to consensus.
However, assertion b) is explicitly needed in the analysis given below, but cannot be treated as a direct
consequence of the results from the literature. \end{remark}

 \subsection{Weak Convergence: the Limit ODE}
Weak convergence of the algorithm (\ref{alg}) will be analyzed under the following additional assumptions:
\par
(A.2) Sequences $\{ A_{n} \}$ and $\{ \xi_{n} \}$ are mutually independent;
\par
(A.3) Let  $ \xi_{n}=(\eta_{n}, \zeta_{n})$; then, $F(Y_{n},\xi_{n})=F_{1}(Y_{n}, \eta_{n})+F_{2}(Y_{n}) \zeta_{n}$ where both $F_{1}( \cdot, \cdot )$ and $F_{2}(\cdot) $ are continuous, $\{ \eta_{n} \}$ is a sequence of bounded in probability random variables and $\{ \zeta_{n} \}$ a random sequence with zero mean and finite fourth moment;
\par
(A.4)  $ \lim_{n-k \to \infty}$ $ E_{k} \{ F_{1}(X, $ $\eta_{n})\}=\bar{F}(X)$ in probability,
where $X=(X^{1}, \cdots, X^{N})$ is a dummy variable, $\bar{F}(X)=(\bar{f}^{1}(X), \cdots, \bar{f}^{N}(X))$ is a continuous function; $E_{k}\{ \zeta_{n} \} \rightarrow 0$ in probability when $n-k \rightarrow \infty$.
\par
(A.5) Let $\bar{D}= E \{ D_{n} \} = {\rm diag} \{ \bar{d}_{1}, \cdots, \bar{d}_{N} \}$. Denote $\bar{f}^{i}(X)|_{X=(x, \cdots, x)}=\bar{f}^{i}(x)$, where $x \in {\bf R}^{p}$ is a dummy variable, $i=1, \ldots, N$; then
the ODE
\begin{equation} \label{ODE}
\dot{x} =  \sum_{i=1}^{N} \bar{\phi}_{i} \bar{d}_{i} \bar{f}^{i}(x)
\end{equation}
has a unique solution for each initial condition.
\par
(A.6) $\{X_{n} \}$ is tight (bounded in probability).
\par
Let $ \tilde{\Phi}(n|k)=\Phi(n|k) \otimes I_{p}$, $ \tilde{\Phi}_{k}=\Phi_{k} \otimes I_{p}$, $\tilde{\Psi}(n|k)=\tilde{\Phi}(n|k+1) \tilde{C}_{k}$ and $\tilde{\Psi}_{k}=\tilde{\Phi}_{k+1} \tilde{C}_{k}$, with $\tilde{C}_{k}$ resulting from (\ref{alg}). Following \cite{ky1}, let $\{ n_{\varepsilon} \}$ be a sequence tending to $\infty$ when $\varepsilon \rightarrow 0$, such that $ \sqrt{\varepsilon} n_{\varepsilon} \rightarrow 0$ and  $\sup_{k} P \{ |\Phi(k+n_{\varepsilon}|k)-\Phi_{k}| \geq
\varepsilon^{2}\} \leq \varepsilon^{2}$. Let $X^{\varepsilon}(t) = X_{n}$ for $t \in [(n-n_{\varepsilon}) \varepsilon, (n-n_{\varepsilon}+1) \varepsilon]$ for $n \geq n_{\varepsilon}$, where $X_{n}$ is defined in (\ref{alg}); also, let
$X_{0}^{\varepsilon}= \tilde{\Phi}(n_{\varepsilon}|0) X_{0}+ $ $\varepsilon \sum_{k=0}^{n_{\varepsilon}-1} \tilde{\Psi}_{k}F(Y_{k},\xi_{k})$.

\begin{theorem} Assume (A.1) -- (A.6). Let $X(0)= \lim_{\varepsilon \to 0 } X^{\varepsilon}_{0}$ $=(x_{0}, \cdots, x_{0})$. Then for both AUC and ACU  $X^{\varepsilon}(t)$ is tight in $D[0, \infty)$ and converges weakly to $X(t)=(x(t), \cdots, x(t))$, $t \in \textbf{R}$,  where $x(t)$
satisfies (\ref{ODE}) with initial condition $x_{0}$ ($D[0, \infty)$ denotes the space of vector valued functions on $[0, \infty)$ which are right continuous
and have left hand limits, with the Skorokhod topology \cite{ky,ky1}). \end{theorem}

{\bf Proof:}  The proof follows the main line of thought of the proof of Theorem 3.1 in \cite{ky1}, where, in fact, a SA algorithm different from (\ref{alg}) has been considered. We will emphasize only the points where the proof differs from the one in \cite{ky1}, due to specific properties of (\ref{alg}).
\par
 After iterating (\ref{alg}) back to the initial condition, we obtain
\begin{align} \label{iterate}
 X_{n+1}=X_{0}^{\varepsilon}
 +\varepsilon \sum_{k=n_{\varepsilon}}^{n} \tilde{\Psi}_{k} F(Y_{k}, \xi_{k})+ \varepsilon \varrho_{n}^{\varepsilon}  + [\tilde{\Phi}(n|0) - \tilde{\Phi}(n_{\varepsilon}|0)] X_{0},
\end{align}
where  $\varrho_{n}^{\varepsilon}= \sum_{k=0}^{n} [\tilde{\Psi}(n|k)-\tilde{\Psi}_{k}] F(Y_{k},\xi_{k})$.
Further, we use (A.2), (A.3) and (A.6) and verify that for both AUC and ACU assertion b) from Lemma~1 implies that
$E \{ |\tilde{\Psi}(n|k)-\tilde{\Psi}_{k}| \} \rightarrow 0$
 geometrically, because $\tilde{\Psi}(n|k)-\tilde{\Psi}_{k}=
 (\tilde{\Phi}(n|k)-\tilde{\Phi}_{k})(D_{k} \otimes I_{p})$ for AUC and $\tilde{\Psi}(n|k)-\tilde{\Psi}_{k}=
 (\tilde{\Phi}(n|k+1)-\tilde{\Phi}_{k+1})(D_{k}^{-} \otimes I_{p})$ for ACU.
  It follows from \cite{ky1} (having in mind that (A.2) -- (A.4) are equivalent to assumptions (C3.1) -- (C3.3) in \cite{ky1}), that $\sup_{\varepsilon, n} E \{ | \varrho_{n}^{\varepsilon}| ^{3} \} < \infty$ and $\sup_{\varepsilon, n \geq n_{\varepsilon}} E \{ | X_{n+1}-X_{n} |^{2} \}/ \varepsilon^{2} < \infty$, implying that $\{ X^{\varepsilon}(t) \} $ is tight in $D[0, \infty)$ and that all limit paths are Lipschitz continuous in $t$.

At the second step, following methodologically \cite{ky,ky1}, we introduce
\begin{align}
  M_{g}(t)=g(X(t))-g(X(0)) +\int_{0}^{t}g'_{X}(X(s)) (\bar{\Phi} \bar{D} \otimes I_{p}) \bar{F}(X(s)) ds 
\end{align}
and show that $M_{g}(t)$ is a martingale for any real valued function $g(\cdot)$ with compact support and continuous second derivatives. As $ M_{g}(t)$ is Lipschitz continuous for both AUC and ACU (because $X(\cdot)$ is Lipschitz continuous), it follows that $M_{g}(t)$ is constant and equal to zero; consequently, $\dot{X} = (\bar{\Phi} \bar{D} \otimes I_{p}) \bar{F}(X)$. Furthermore, as the rows of $\bar{\Phi} \bar{D} $ are equal, all the $p$-vector components of $X(t)$ must be equal, implying further that  $X^{\varepsilon}(t)$
converges weakly to $X(t)=(x(t), \cdots, x(t))$, where $x(t)$ satisfies the ODE (\ref{ODE}). \par In order to show the martingale property, the main point is to demonstrate that
 \begin{equation} \label{novo}
 B^{\varepsilon}(u)  \stackrel{\varepsilon \to 0}{\rightarrow} g'_{X}(X(u)) (\bar{\Phi} \bar{D} \otimes I_{p}) \bar{F}(X(u)),
 \end{equation}
for almost all $u$, where
$B^{\varepsilon}(t)=$ $ g'_{X}(X^{\varepsilon}_{l+n_{\varepsilon}})$ $E_{l+n_{\varepsilon}} \{ \tilde{\Psi}_{l+n_{\varepsilon}} F(Y_{l+n_{\varepsilon}},$ $ \xi_{l+n_{\varepsilon}}) \} $, $l \varepsilon \leq t$ $ < (l+1) \varepsilon$ (compare with \cite{ky1}, proof of Theorem~3.1, Eqs. (3.6) -- (3.10)).

Due to the independence assumption (A.2), we have for AUC that
\begin{eqnarray}
 & E \{ \tilde{\Psi}_{k} \} = E\{ \tilde{\Phi}_{k} ( D_{k} \otimes I_{p}) \}=\bar{\Phi} E\{ A_{k} D_{k}\} \otimes  I_{p} \\
 & =\bar{\Phi} \sum_{l} A^{[l]}D^{[l]} \pi_{l} \otimes I_{p}=\bar{\Phi}(\bar{A}-\sum_{l}A^{[l]} D^{[l]-} \pi_{l} ) \otimes I_{p}, \nonumber
 \end{eqnarray}
  where $D^{[l]}$ are realizations of $D_{n}$ corresponding to the realizations $A^{[l]}$ of $A_{n}$ and
  $D^{[l]-}=I-D^{[l]}$. As $\bar{\Phi} \bar{A}=\bar{\Phi}$ and  $A^{[l]} D^{[l]-}=D^{[l]-}$, because of the specific structure of $A^{[l]}$, one concludes that
\begin{equation}
E \{ \tilde{\Psi}_{k} \}= \bar{\Phi}(\bar{A}-\bar{D}^{-}) \otimes I_{p}=\bar{\Phi} \bar{D} \otimes I_{p},
\end{equation}
 where $\bar{D}^{-}= \sum_{l}D^{[l]-} \pi_{l}$ is composed of the node probabilities of not updating the state at a given instant $n$.
\par
For ACU, $E \{ \tilde{\Psi}_{k} \}=\bar{\Phi} \bar{D} \otimes I_{p}$, since $A_{k+1}$ and $D_{k}$ are independent.
\par
On the other hand, the term $E_{l+n_{\varepsilon}} \{F(X_{k}, \xi_{k}) \}$ converges to $\bar{F}(X(u))$ when $\varepsilon \to 0$, because of the convergence of $X^{\varepsilon}(\cdot)$ to $X(\cdot)$ and assumptions (A.3) and (A.4). Therefore, (\ref{novo}) holds, which concludes the proof. \hspace*{\fill}\QED

\subsection{Asymptotics for Large $t$}
 (A.7) Let ODE (\ref{ODE}) has a unique, in the sense of Lyapunov, stable point $x^{*}$ which is globally attracting.
 \par
\begin{theorem}
Let the conditions of Theorem 1 and (A.7) be satisfied. Then $X^{\varepsilon}(t)$ converges weakly to $X^{*}=(x^{*}, \cdots, x^{*})$ as $t \to \infty$ and $\varepsilon \to 0$. \end{theorem}

{\bf Proof:} The proof follows from Theorem~1; it is analogous to Theorem~5.1 in \cite{ky1}.  \hspace*{\fill}\QED

\subsection{Convergence Rate}
Sharper bounds on the asymptotic normalized error $U_{n}^{\varepsilon}=
\frac{X_{n}^\varepsilon-X^{*}}{\sqrt{\varepsilon}}$ can be obtained using \cite{ky1}, after assuming that $U_{n}^{\varepsilon}$ is tight and that, additionally, $\bar{F}(X^{*}) =0$. Define $W_{n}^{\varepsilon}= \sqrt{\varepsilon} \sum_{k=M+n_{\varepsilon}+1}^{n} \tilde{\Psi}_{k} F(Y_{k}, \xi_{k})$ where $M$ is large enough. Because of the lack of space, we will only indicate that it is possible to demonstrate under additional assumptions (including independence of $\xi^{i}_{n}$, $i=1, \ldots, N$), that for both AUC and ACU $U_{n}^{\varepsilon}$ and $  W_{n}^{\varepsilon}$ converge weakly to $U(\cdot)=(u(t), \cdots, u(t))$  and $W(\cdot)=(w(t), \cdots, w(t))$, respectively, where $u(t)$ and $w(t)$ satisfy the following linear stochastic differential equation
\begin{equation} \label{sde}
du=J u dt +dw,
\end{equation}
in which $J=((\bar{\Phi}\bar{D} \otimes I_{p}) \bar{F}(x^{*}))_{x}$ is the Jacobian of $(\bar{\Phi}\bar{D} \otimes I_{p})\bar{F}(x)$ at $x=x^{*}$ and $w(\cdot)$ is a $p$-dimensional Wiener process, with covariance
\begin{equation} \label{cov}
{\rm cov} w(1)= \sum_{i=1}^{N} g_{i} R_{i},
\end{equation}
 where $R_{i}= {\rm cov} F^{i}(x^{*}, \xi_{k}^{i})$ and $g_{i} = E\{ \phi_{i}(k)^{2} d_{i}(k)^{2} \}$ for any time instance $k$. We show in Subsection~IV.B, Remark~8, how this theoretical result can be efficiently used in practice for improving convergence rate of the algorithm.

\begin{remark} The above analysis can be extended to the case when $\varepsilon$ is replaced by a time varying sequence $\{ \varepsilon_{n} \}$, such that $\varepsilon_{n} > 0$,  $\varepsilon_{n} \to 0$ and $\sum \varepsilon_{n} = \infty$, as it is common in SA procedures \cite{ky,hfchen}. Starting from the assumptions of Theorem~1, one can readily apply the standard methodology from \cite{ky} and obtain that \emph{the same ODE as above} characterizes the limit paths of the estimates. The w.p.1 convergence results can be derived assuming that $\sum \varepsilon^{2}_{n} < \infty$; details related to tapering step-sizes are out of the scope of this paper (some simulations are given in Section~V).  \end{remark}

\begin{remark} In \cite{ned} an asynchronous algorithm of ACU type for broadcast-based convex minimization of sum of
 local criteria has been analyzed assuming bidirectional links. In \cite{bfh} and \cite{mbf}, convergence of a DSA algorithm of AUC type with tapering
 step size is proved under the additional assumption that $ \rho (E \{ A_{n}^{T}(I-\mathbf{11}^{T}/pN) A_{n} \} )< 1$
 ($\rho(A)$ indicates the spectral radius of a matrix $A$), which is not required above for weak convergence.   \end{remark}

\section{Network Design}
\subsection{Communication Scheme}
The analysis in Section~III.A opens up an important possibility of \emph{designing the network} \emph{in accordance with the desired asymptotic behavior at consensus} of the communication scheme taken apart form the SA iterations. Consider the following problem: \par Let $\bar{\phi}=(\bar{\phi}_{1}, \cdots, \bar{\phi}_{N})^{T}$, ($0 \leq \bar{\phi}_{i} < 1$, $\sum_{j}\bar{\phi}_{j}=1$), be given; for a given structure of digraph ${\cal G}$ satisfying (A.1), find parameters $p_{i} > 0$, $i=1, \ldots, N$, $\sum_{i}p_{i}=1$, and $\beta_{ij}$, ($0 < \beta_{ij} < 1$),  $i,j=1, \ldots, N$, such that $\lim_{n \to \infty} \bar{A}^{n}=\left[ \begin{BMAT}{ccc}{c} \bar{\phi}^{T}  & \cdots & \bar{\phi}^{T}  \end{BMAT} \right]^{T}$.
\par
Formally, we have to solve the standard equation $\bar{\phi} \bar{A} = \bar{\phi}$ for the unknown parameters $p_{i}$ and $ \beta_{ij}$ under the given constraints, $i,j=1, \ldots, N$. Obviously, for $\bar{\phi}= \mathbf{1}^{T}$ any doubly stochastic matrix $\bar{A}$ satisfying the given constraints is a solution. The same idea has been the starting point for constructing a recursive distributed matrix scaling algorithm in the deterministic time-invariant \cite{dgh} and time-varying context \cite{neol}; however, these approaches cannot be applied in the case of broadcast gossip scheme. Our focus is on an \emph{off-line} approach, in which the whole network is designed before being used for distributed recursive computations. We assume first, for the sake of exposition clarity, that ${\cal J}_{i}(n)= {\cal N}_{i}^{o}$. If $A^{[i]}$ is the realization of $A_{n}$  corresponding to a tick of the $i$-th clock, we have then the basic relation $\bar{\phi} \sum_{i} A^{[i]} p_{i} = \bar{\phi}$, connecting the unknown parameters with the desired asymptotic vector $\bar{\phi}$. We propose in the sequel two algorithms, both providing efficient solutions to the problem: Algorithm A) - parameters $\beta_{ij}$ are fixed, probabilities $p_{i}$ are unknown; Algorithm B) - probabilities $p_{i}$ are fixed, parameters $\beta_{ij}$ are unknown, $i,j=1, \ldots, N$.

\begin{theorem}{(\textit{Algorithm~A})} For any admissible values of $\bar{\phi}_{i}$  and $\beta_{ij}$, $i,j=1, \ldots, N$,
 \begin{equation}
 p=  (\bar{L}^{p})^{-1} \left( \begin{BMAT}{cccc}{c} 0 & \cdots & 0  & 1 \end{BMAT} \right)^{T}
 \end{equation}
 is the unique solution of $\bar{\phi} \bar{A} = \bar{\phi}$ with respect to $p = (p_{1},  \cdots, p_{N})$, where $\bar{L}^{p}=
 \left[ \begin{BMAT}{c}{c.c} \bar{L}^{A'} \\ \mathbf{1}^{T} \end{BMAT} \right]$, while $\bar{L}^{A'}$ is composed of any set of $N-1$ rows of
\[\bar{L}^{A}=
 \left[ \begin{BMAT}{c.c.c.c}{c.c.c.c} \sum_{j,j \neq 1}  \bar{\phi}_{j} \beta_{1j} & -\bar{\phi}_{1} \beta_{21} & & -\bar{\phi}_{1} \beta_{N1} \\ -\bar{\phi}_{2} \beta_{12} & \sum_{j,j \neq 2}  \bar{\phi}_{j} \beta_{2j} & \cdots & \vdots \\ \cdots & & \cdots & -\bar{\phi}_{N-1} \beta_{N,N-1} \\ -\bar{\phi}_{N} \beta_{1N} & \cdots & &  \sum_{j,j \neq N}  \bar{\phi}_{j} \beta_{Nj} \end{BMAT} \right]. \] \end{theorem}

{\bf Proof:} Relation $\bar{\phi} \sum_{i} A^{[i]} p_{i} = \bar{\phi}$ can be written as $\bar{L}^{A}p=0$. As $(\bar{L}^{A})^{T}$ is in the form of a \emph{weighted Laplacian} of  ${\cal G}$, it follows that ${ \rm rank} \{ \bar{L}^{A'} \}=N-1$  under (A.1); the last row of $\bar{L}^{p}$ follows from  $\sum_{i=1}^{N} p_{i}=1$, and is linearly independent of the rows of $\bar{L}^{A'}$.  It is easy to check that $0 < p_{i} < 1$, $i=1, \ldots, N$.  \hspace*{\fill}\QED

\begin{theorem}{(\textit{Algorithm~B})} For any fixed admissible values of $\bar{\phi}_{i}$ and $p_{i}$, $i=1, \ldots, N$, all solutions of
\begin{equation}
\bar{L}^{B} \beta =0
\end{equation}
with respect to $\beta=(\beta_{1}, \cdots, \beta_{N} )$ ($\beta_{ij}=\beta_{i}$ for all $i,j=1, \ldots, N$), represent admissible solutions of $\bar{\phi} \bar{A} = \bar{\phi}$, where
\[ \bar{L}^{B}=  \left[ \begin{BMAT}{c.c.c.c}{c.c.c.c}  p_{1} \sum_{j,j \neq 1} \bar{\phi}_{j}  &  - p_{2} \bar{\phi}_{1} & \cdots & - p_{N} \bar{\phi}_{1} \\ - p_{1} \bar{\phi}_{2} & p_{2} \sum_{j, j \neq 2} \bar{\phi}_{j}
  & \cdots & - p_{N} \bar{\phi}_{2} \\  & & \cdots &  \\ - p_{1} \bar{\phi}_{N}  & \cdots & &  p_{N} \sum_{j, j \neq N}  \bar{\phi}_{j}  \end{BMAT} \right], \]
satisfying ${\rm rank} \{\bar{L}^{B} \}=N-1 $. \end{theorem}

{\bf Proof:} $\bar{\phi}\bar{A} = \bar{\phi}$ implies $\bar{L}^{B} \beta =0$;  $(\bar{L}^{B})^{T}$ has the form of \emph{weighted Laplacian} of  ${\cal G}$, so that ${\rm rank} \{ \bar{L}^{B} \} =N-1$ under (A.1). It is easy to conclude that there is an infinite number of solutions for $\beta$ satisfying $0 < \beta_{i} < 1$, $i=1, \ldots, N$. In general, when $\beta_{ij}$ are different for different $j$, we have additional degrees of freedom. \hspace*{\fill}\QED

In the general case of communication outages (when ${\cal J}_{i}(n) \neq {\cal N}_{i}^{o}$), a similar analysis can be done, using transmission probabilities $p_{ij}$.

\begin{remark}
  According to the above results, consensus averaging, characterized by $\bar{\phi}=\frac{1}{N} {\bf 1}^{T}$, can be achieved \emph{for any given network structure} by choosing either of the two proposed algorithms.  \end{remark}

 \subsection{Main DSA Algorithm}
The aim of the network design for (\ref{alg}) is to provide the desired asymptotic behavior defined by $\dot{x}=0$ in ODE (\ref{ODE}). The basic tools are already presented in Subsection~IV.A. However, in addition, ODE (\ref{ODE}) incorporates the node updating probabilities $\bar{d}_{i}$, $i=1, \ldots, N$ (obtainable directly from $p_{i}$ and $p_{ij}$). Assume that the designer's goal is to obtain a predefined set of weights $w_{i}= \bar{\phi}_{i} \bar{d}_{i}$, $w_{i} > 0$, $i=1, \ldots, N$. Then, we have the following design procedure: \par 1) adopt a set of transmission probabilities $p_{i} > 0$ and define $\bar{d}_{i}$, $i=1, \ldots, N$, and \par 2) apply Algorithm B from IV.B by choosing such $\beta_{ij}$ that the set $\bar{\phi}_{k}=w_{k}/\bar{d}_{k}$, $k=1, \ldots, N$ is obtained at consensus (after appropriate normalization).

\begin{remark} If, for example, the proposed DSA algorithms are aimed at distributed minimization of a given criterion $J=\sum_{i=1}^{N} w_{i} E \{  J_{i}(x, \xi^{i}_{n}) \} $, $w_{i} > 0$, then $w_{i}$ are taken as a set of predefined weights; in this case $F^{i}(X^{i}, \xi^{i}_{n})$ in (\ref{auc1}) and (\ref{acu2}) are noisy gradients or pseudo-gradients of $J_{i}(x, \xi^{i}_{n})$ at $x=X^{i}$ (see \emph{e.g.} \cite{ned,no,lo}). Obviously, the case of equal weights (consensus averaging) can be obtained \emph{for any given network topology} satisfying (A.1) (much more restrictive assumptions are adopted in \cite{nop,rnv,ned,lo}). \end{remark}
 \par
\begin{remark} Notice that an additional degree of freedom can be obtained by choosing different gains $\varepsilon^{i}=v^{i} \varepsilon$ for different agents, since we have then in (\ref{ODE}) coefficients of the form $w_{i} = $ $v_{i}\bar{\phi}_{i}\bar{d}_{i}$. \end{remark}
\par
\begin{remark} Analogous conclusions as above hold for tapering step-sizes, since the same ODE holds in this case. However, choosing the same gain $\varepsilon_{n}$ for all the agents leads to network centralization through the need for a centralized clock determining $n$. This problem can be overcome by \emph{asynchronous strategies} in which the step-size of the $i$-th agent is given by  $\varepsilon_{i}({\Gamma_{i}(n)})$, where $\Gamma_{i}(n)$ is the number of updates of the agent $i$ up to the instant $n$ (see, \textit{e.g.}, \cite{bor,ned}). \end{remark}

\begin{remark}{\textit{(Convergence Rate Optimization)}} A substantially different possibility for network design follows from the analysis of the rate of convergence in Section~III.D; relations (\ref{sde}) and (\ref{cov}) can be used to reduce the stationary covariance of $u$ in (\ref{sde}) by minimizing  the upper bound of  $\| {\rm cov} w(1) \|$. Since $E\{ \phi_{i}(k)^{2} d_{i}(k)^{2} \}$ is not easy to calculate, one can formulate the following practical criterion to be minimized: $\sum_{i=1}^{N}  \bar{\phi}_{i}^{2} \bar{d}_{i}^{2}  \| R_{i} \|$. The optimal value of $\bar{\phi}_{i} $ satisfying $\sum_{i} \bar{\phi}_{i}=1$  is $ \bar{\phi}_{i}^{*} $ $=
\bar{d}_{i}^{-2} \|R_{i}\|^{-1} / \sum_{k} \bar{d}_{k}^{-2} \| R_{k} \|^{-1}$. Concrete network parameters can now be found by using one of the two proposed methodologies in IV.A, starting from the obtained $ \bar{\phi}_{i}^{*} $ $i=1, \ldots, N$; notice that the assumption $E\{ F(X^{*}, \xi_{n})\}=0$ implies that the convergence point at consensus does not depend on the network parameters, see (\ref{ODE}). An example is given in Section~V. \end{remark}

\section{Simulation Results}

\emph{Example 1:} A sensor network with $N=10$ nodes represented by a digraph has been simulated.
Algorithm AUC has been applied to distributed parameter estimation with $F^{i}(X_{n}^{i}, \xi_{n}^{i})=
\mu_{n}^{i}-X_{n}^{i}$, where $\mu_{n}^{i}$ is a random variable $\mu_{n}^{i} \sim {\cal N}(m_{i}, \sigma_{i}^{2})$,
$m_{i}$ and $\sigma_{i}$ being randomly selected from the intervals [3,7] and [1,5], respectively.
The network parameters have been chosen in such a way as to achieve equal weights in ODE (\ref{ODE}), \emph{i.e.},
 $\bar{\phi}_{i} \bar{d}_{i} = \bar{\phi}_{j} \bar{d}_{j}$, $i,j=1, \ldots, N$. At the first step, probabilities
 $p_{i}$ have been chosen, determining implicitly the values of $\bar{d}_{i}$, $i=1, \ldots, N$. At the second step, the values  $\bar{\phi}_{i}=\bar{d}_{i}^{-1}/\sum_{j=1}^{N} \bar{d}_{j}^{-1}$ have been used to define parameters $\beta_{i}$, $i=1, \ldots, N$, according to Algorithm B in Subsection~IV.A. One degree of freedom in  $\bar{A}^{EL} \beta =0$ has been utilized to set $\max_{i} \beta_{i}$ at its maximal value close to one; in such a way, all $\beta_{i}$ take their own maximal values, increasing convergence rate. Fig.~1 gives an illustration of the algorithm behavior; the straight line defines the asymptotic parameter value at consensus. Algorithm ACU has the same asymptotic ODE and very similar behavior. However, AUC has been found to be slightly superior in practice, especially in large networks with high connectedness, because of explicit noise averaging done by convexification.
\begin{figure}
\begin{center}
\includegraphics[width=6.5cm]{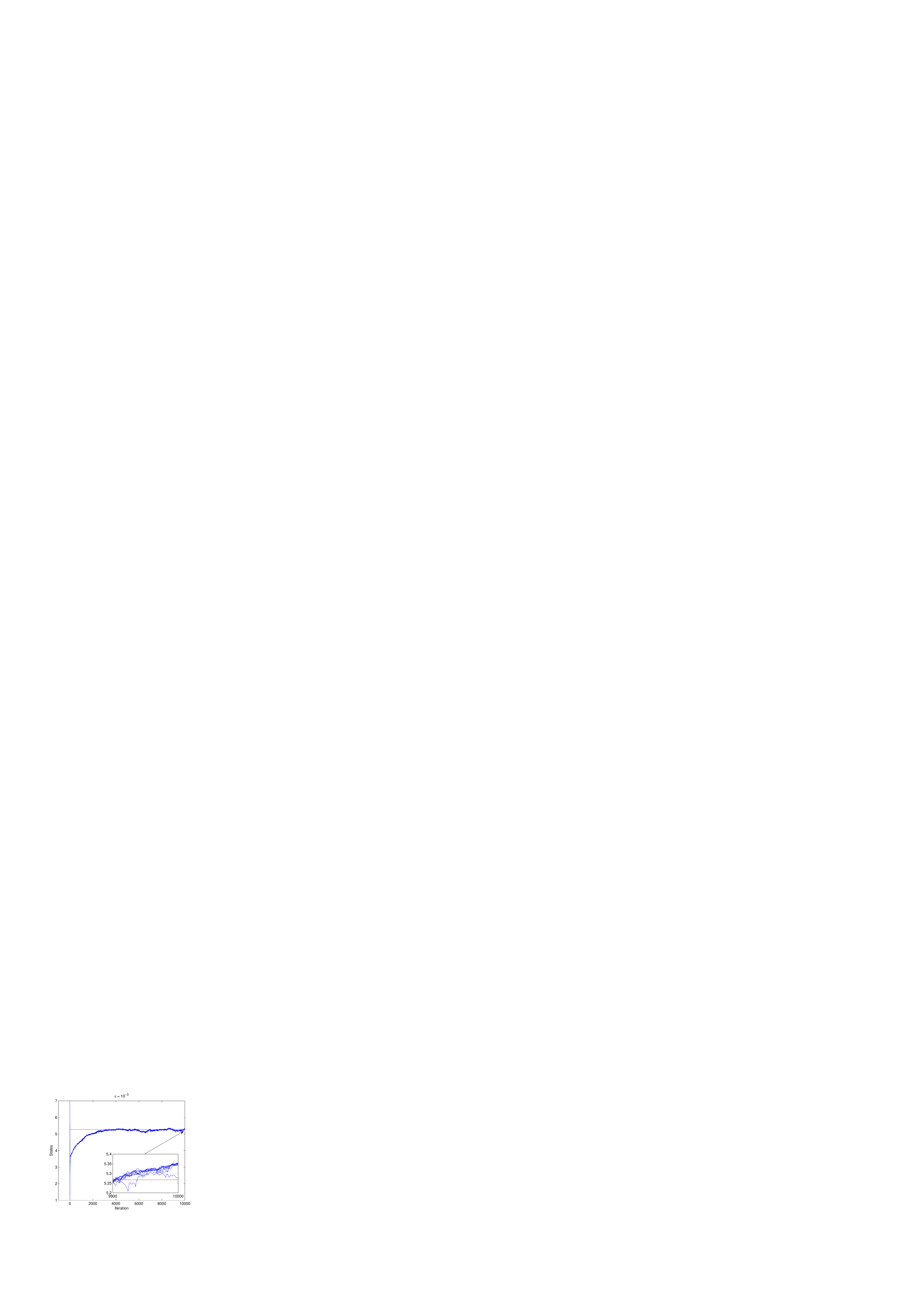}    
\end{center}
\caption{Estimates of all nodes for AUC}
\end{figure}

\begin{figure}
\begin{center}
\includegraphics[width=6.5cm]{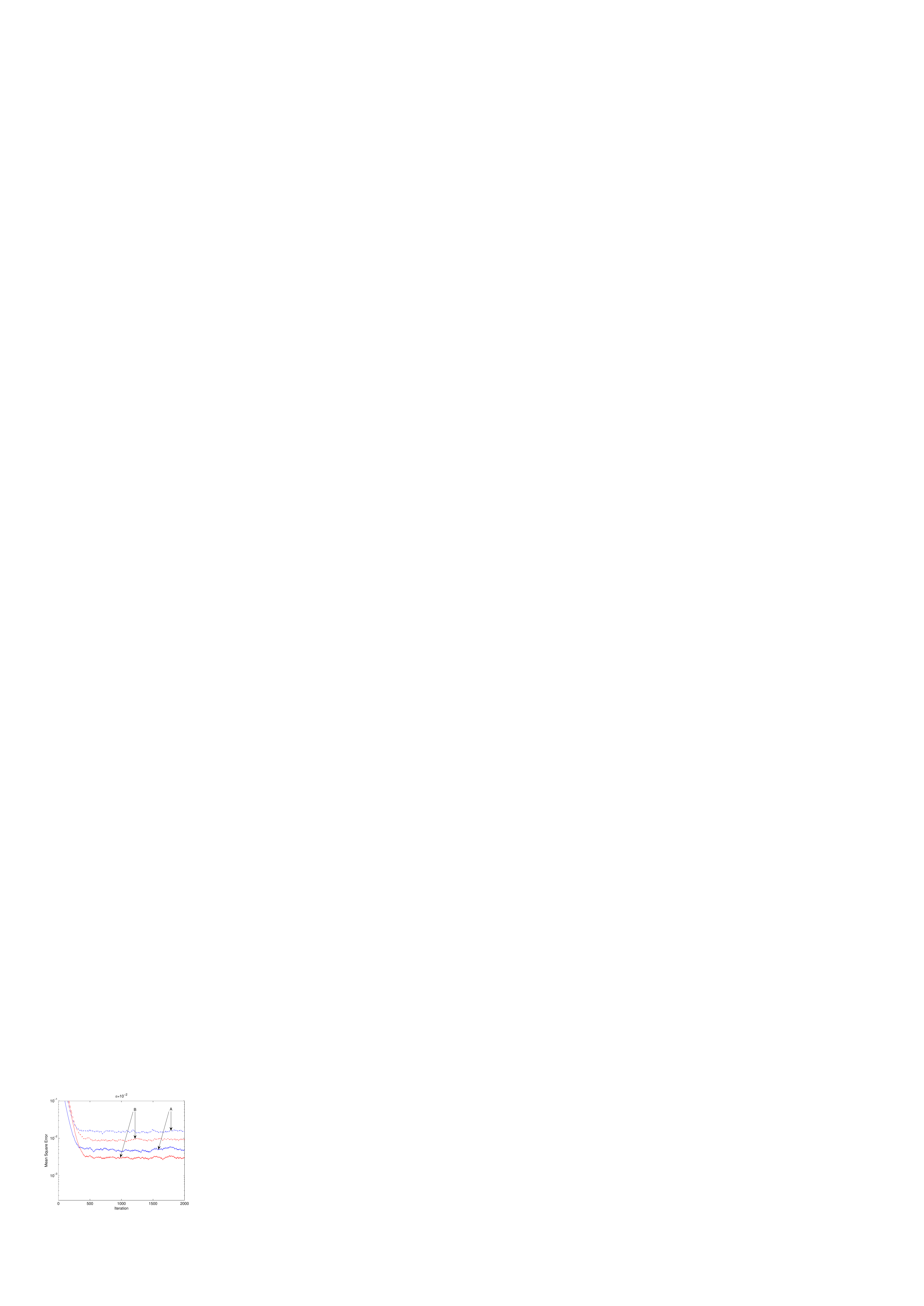}    
\end{center}
\caption{Mean-square error for AUC, constant step-size: constant consensus matrices (solid lines), gossip (dotted lines); equal weights (A), optimized network parameters (B)}
\end{figure}

\begin{figure}
\begin{center}
\includegraphics[width=6.5cm]{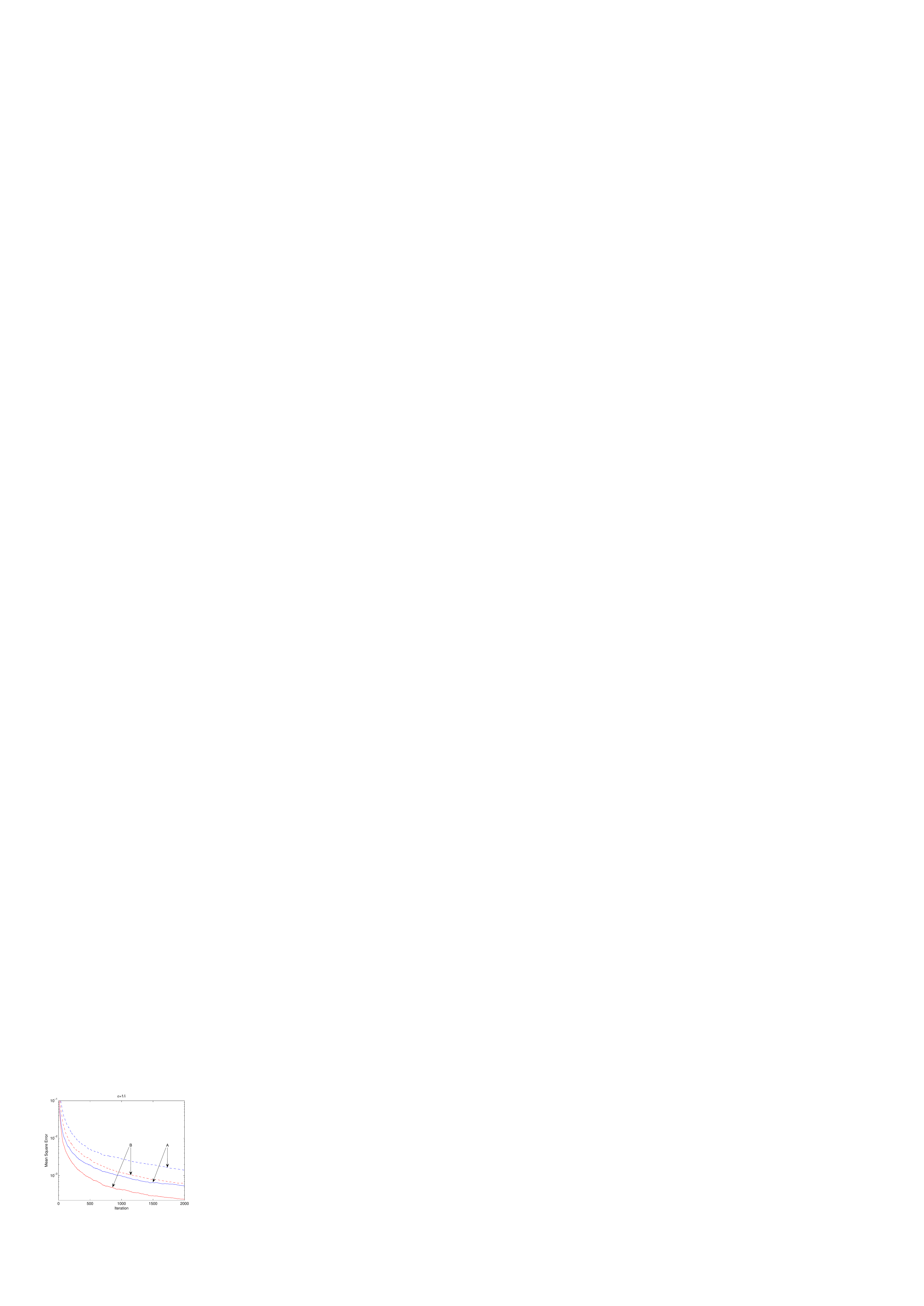}    
\end{center}
\caption{Mean-square error for AUC, tapering step-size $\varepsilon_{n}=1/n$: constant consensus matrices (solid lines), gossip (dotted lines); equal weights (A), optimized network parameters (B)}
\end{figure}
\par
\emph{Example 2:} The same network, but with fixed $m_{i}=5$, $i=1, \ldots, N$, has been used to check network optimization based on the asymptotic SDE (\ref{sde}), described in Remark~8. Since in this case $\bar{f}^{i}(x^{*})$ are equal for all $i$, $\bar{\phi}_{i}$ and $\bar{d}_{i}$ do not influence the parameter value at convergence. Using the proposed method, we start from (\ref{cov}) and choose $\bar{\phi}_{i}=$
 $\bar{\phi}_{i}^{*} = \bar{d}_{i}^{-2}\sigma_{i}^{-2}/ \sum_{j} \bar{d}_{j}^{-2}\sigma_{j}^{-2}$, $i=1, \ldots, N$. Fig~~2 shows that such an approximate optimization provides a substantial advantage with respect to the case of equal values of $\bar{\phi}_{i}^{*}$ (dotted lines). In addition, in order to make clear the overall performance of the proposed algorithm, the case $A_{n}=\bar{A}$ (fixed deterministic communication network) is represented by solid lines in Fig.~2.

In order to show that the weak convergence results can be directly extended to the case of tapering step-size, in Fig.~3 curves analogous to those in Fig.~2 are presented, but with $\varepsilon_{n}=1/n$.

\section{Conclusion}
In this paper distributed stochastic approximation algorithms based on asynchronous broadcast gossip on networks represented by digraphs have been considered. Convergence w.p.1 of the main gossip scheme has been proved. Weak convergence results have then been derived for the main algorithm, resulting in the formulation of the associated limit ODE connecting convergence points with the network parameters and probabilities of the nodes to communicate and update their estimates. Two network design schemes have been proposed for ensuring the desired asymptotic behavior of the gossip scheme itself, on one side, and of the parameter estimates in DSA algorithms, on the other. Using a limit stochastic differential equation for the normalized asymptotic error, a method for convergence rate improvement has also been proposed.

\bibliography{formifac}
\end{document}